\documentclass[12pt]{iopart}
\usepackage[utf8]{inputenc}
\usepackage[T1]{fontenc}
\usepackage{graphicx}
\usepackage{babel}
\usepackage{iopams}
\usepackage{hyperref}
\usepackage{xcolor}

\begin{document}
	
	\title [Stress dynamics in chain of springs and masses]{Stretched-exponential stress dynamics in chain of springs and masses model of crystals: analytical results and MD simulations.}  
	
	\author{Zbigniew Kozio{\l}}
	
	\address{National Center for Nuclear Research, Materials Research Laboratory\\ul. Andrzeja So{\l}tana 7, 05-400 Otwock-{\'S}wierk, Poland}
	\ead{zbigniew.koziol@ncbj.gov.pl}
	
	\vspace{10pt}
	
	\begin{abstract}
		The model of a chain of springs and masses (CSM), originating from the works of Schr\"odinger (1914), and Pater (1974), is found suitable as an analytical description of the dynamics of layers in orientated FCC crystals. An analytical extension of that model has been developed for the case of linear-in-time ramp pressure applied to a sample surface. Examples are provided of molecular dynamics (MD) simulations, confirming the usefulness of the model in the description of dynamic effects on steel 310S under pressure. Qualitatively the same results have been obtained by us for several other medium-entropy alloys CoNiCr (with EAM and MEAM inter-atomic potentials), and CoNiV (with MEAM potential). For studies of dynamic stability on large sizes of samples and for long times, the proposed earlier table-style harmonic interlayer potential has been used. The results of MD simulations suggest that the dynamics of the CSM model of perfectly ordered matter is described by stretched-exponential time functions, and it is characterised by simple scaling relations in time and size of the sample.
	\end{abstract}
	
	\vspace{2pc}
	
	\maketitle
	
	\section{Introduction.}
	\label{Introduction}
	 
This work may be considered as a continuation and an expansion of our recently reported results on the analysis of dynamics of stress in anharmonic crystals \cite{Koziol0}. Presently, we concentrate on the analysis of effects occurring in materials governed by harmonic inter-atomic (and interlayer) potentials at large sample sizes and at long times. We use a theoretical framework of a model of a chain of springs and masses (CSM), where the movement of masses connected by elastic massless springs is investigated in 1 dimension.

The model itself has been analysed for over a century, for instance, by Erwin Schr\"odinger in 1914 \cite{Erwin}, \cite{Muhlich}. 

Numerous variants of the CSM model have been used often to study aspects of non-linearities, chaos, and localization/de-localization of lattice vibrations, phonon density of states \cite{Nucera,Barreto,Pankov,Kashchenko,Santos,Parmley}. A very close mathematical problem of an electrical signal transmission in a ladder of periodically arranged capacitive, inductive, and resistive elements \cite{Greco} was shown to be related to Fibonacci wave functions \cite{Hissem}, \cite{Hissem2}. It was however solved numerically only, and for a limited number of ladder nodes.

The approach based on the model of masses connected by springs was found useful in the case of explaining, for instance, Raman spectra in a few layers of graphene \cite{Lui}, \cite{Tan}, and in our analysis of Van der Waals interaction in graphene \cite{Koziol}.

Today's computational possibilities go far beyond those present when some of the cited works were created. This is not only due to computer hardware development but also the availability of advanced simulation software, like \texttt{lammps} \cite{LAMMPS}. 

Our interest in the CSM model grew gradually during the last 2-3 years \cite{Koziol2}, \cite{Koziol0}, as we realised that the model is suitable for describing the dynamics of penetration of stress waves into crystals, while at the same time it has become clear to us that no a broadly acceptable model of that phenomenon is used by the community of researchers studying, with the use of MD simulations, for instance, retarded in time effects related to the  dislocations movement. In a typical simulation setup, the symmetry of samples used and forces applied is such that the movement of atoms occurs basically in one direction only, i.e., that we can consider the movement of crystallographic planes as a whole, while forces acting in the other two perpendicular directions compensate to zero, at least in the first approximation. 

Another reason for enforcing our focus on the model was discovering an old paper by de Pater (1974) \cite{Pater}. Our first analysis of MD simulation results using exact analytical equations of de Pater confirmed the correctness and usefulness of that theory.

For these reasons, first, we present examples of using equations derived by de Pater in the analysis of results obtained on samples of steel 310S. 
We provide an extension of Pater's theory, making it useful not only in studies of an impulse response (Heaviside-type time-dependence of forces applied) but also to the case often used as well when the force exerted on a sample grows linearly with time.

For studies of the stability of de Pater's solutions, it was necessary to design samples long in one direction and to use a harmonic interlayer potential. 

In section \ref{Samples}, we describe the simulation setup, creating samples of steel 310S, and samples with a large number of layers in the Y-direction for long-time simulations using harmonic interlayer potential. 

In the next section \ref{PaterAnalysis}, we provide examples of using the exact analytical solutions of de Pater in the analysis of simulation results. It is shown how to reconstruct the dependence on time of quantities like surface pressure, displacement of layers and their velocity, and the virial stress within the entire sample volume. For that we also use our analytical extension of the model of de Pater, derived in \ref{AppendixA}.

We consider the content of section \ref{DynamicsProperties} as the most innovative and important. There, we analysed for the first time the dynamical properties of samples with a large number of layers, and for long times. There are a few astonishing, surprising observations carried out. Some of these are as follows: 
1) When attempting to determine the speed of sound from time-moments when a certain critical value of stress (or a critical value of velocity $v_c$ of a crystallographic plane) is reached for any given layer, it is found that the sound speed must, in that method, depend on time. When, however, $v_c$ equal to $v_0/3$ is used, where $v_0$ is the value of velocity at an asymptotically long time, the speed of propagation of a stress wave does not depend on time.
2) The effect 1) is related to the change with time of the slope of stress profiles of propagating wave. That slope is found to decrease with distance from the surface of a sample (or with time) as a power function of distance/time.
3) Effects 1) and 2) led us to find that a time-scaling property is present in the dynamics of layers.
4) Frequency of oscillations in time of quantities like stress or velocity of layers increases with time/distance and asymptotically reaches a constant value. The time dependence of that frequency is well described by a stretched-exponential function.
5) Amplitude of these oscillations decreases with time, and it can be approximated well either by a power-law or a stretched-exponential relation on time.

	\section{Samples and inter-atomic potentials.}
	\label{Samples}
	
	The description of dynamic processes covered in this work is valid for several FCC materials studied by us: steel 310S and medium entropy alloys 
	CoNiCr (with EAM and MEAM inter-atomic potentials), and CoNiV (MEAM potential). The examples shown here will, however, be limited to those obtained for steel only. 
	
	For the creation of samples of steel 310S (FCC structure), we used either \texttt{lammps} \cite{LAMMPS} or Atomsk \cite{Atomsk}. The sample
	 has an orientation, as usually used in simulations of dislocations \cite{Osetsky}, with the X direction along [$1\bar{1}0$] crystallographic direction, while Y has [$111$] orientation, and Z has [$\bar{1}\bar{1}2$], as visualised in \cite{Koziol0}. In the case of this particular orientation, the X-Z plane becomes the gliding plane for the dislocation movement since the distance between planes in the Y direction is the largest possible, resulting in the weakest interaction between planes. For that reason, this orientation is the most suitable for studies of the CSM model, where masses represent crystallographic layers. It follows also from the symmetry considerations that when force is applied in the X-direction, no force component arises in the Y and Z directions; crystallographic planes move in the X-direction as a whole.
	
	We choose the EAM potentials of Bonny et al. \cite{Bonny} and Artur \cite{Artur} as these that reproduce correctly the basic physical properties of the material and are the most efficient computationally. In the case of steel 310S, we use the chemical composition of atoms as in the specification of that material, as described in \cite{Koziol2}, with FCC lattice constant $a$=3.56 {\AA}. The size of steel samples is 120 unit cells in the orientated X-direction, 20 units in Z, and 16 layers in the Y-direction (with 4800 atoms in every layer). The sample has been annealed, and its temperature is stabilised at 50 K during simulations. We must be aware, though, that contrary to the common belief, stabilising temperature does have an influence on the dynamics of stress waves, disturbing the accuracy of simulations. In our case, though, the effect was negligible. An important detail of the simulation is that the number of atoms in the upper layer is the same as in all other layers. To achieve that, after the sample annealing, one or more surface layers on both sides in the Y-direction are removed.
	
	The upper region is defined as rigid, and the force is applied to atoms in that region in the X-direction. The force \textbf{F} applied to atoms must have a value such that when summed for  all atoms in the \textbf{upper} region and divided by the surface area in the X-Z direction it will result in the desired value of pressure $P^{\rm 0}_{\rm{xy}}$.
	
On the bottom surface (\textbf{lower} region), conditions are imposed with zero forces and velocities of atoms. 

	We find no reason, by testing, why the width of the \textbf{upper} and \textbf{lower} regions may not be made as small as the distance between atomic layers in the Y-direction, i.e., that these regions contain only one layer of atoms.
	In fact, it is the requirement that these regions (in particular the \textbf{upper} one, where force is applied) contain one only layer of atoms. Otherwise, the results of simulations will become untestable by our CSM model: the dynamics depends on the weight (number of atoms) in every layer, including the upper one.
	
	In the case of studies of the dynamics when a very large number of layers N is used, our samples were retained at temperature T=0 K. In this case, the size of samples in X- and Z-directions was 2 orientated unit cells, and that resulted in 8 atoms in every crystallographic layer. 
	
	Statistical averaging of physical quantities along the X direction was done by using Perl scripts. That step, i.e., the proper averaging of data within well-defined layers of atoms is one of the most crucial ones in the entire data analysis. We provide for downloading, as an example, our Perl code \cite{MendeleyData2}, with sample data for analysis.
	
When performing simulations on long samples with harmonic potential, 
for the convenience of numerical computation, we use the following formula for the table-style potential used in \texttt{lammps}:

\begin{equation}
	E_p(r)=\epsilon_0\left(-1+\left|(r-r_0)\right|^{n}\right),
	\label{eq:potential}
\end{equation}

where $r$ and $r_0$ are in {\AA}, $r_0=a/\sqrt{2}$, with $a$ being the FCC lattice constant, and $n$ could be any positive value larger than 1. That potential is 0 at $r-r_0=1$ [{\AA}] and it has value $-\epsilon _0$ at $r=r_0$, for any $n$.

That kind of inter-atomic potential results in the inter-layer potential $E_p$ for single atom well approximated (with respect to the potential minimum) by the following formula \cite{Koziol0}:

\begin{equation}
	E_p(x) -E_p(0) \approx 2\epsilon_0 \cdot \left|\frac{x}{2}\right|^n.
	\label{eq:potentialApprox}
\end{equation}

	One ought to realise that while the inter-atomic potential introduced between atoms is  given by $x^n$ function, the inter-layer potential differs from that form, though not significantly. Our first interpretation of simulation results on long samples was that very small anharmonic contributions to the inter-layer potential cause undesirable dynamic effects in simulation results. For that reason, we constructed an improved version of the inter-atomic potential where contributions of higher order in $n$ have been mostly eliminated. As shown in \cite{MendeleyData0}, when n=2 we ought to use the following
	formula instead of \ref{eq:potential}:

\begin{equation}
	E_p(r)=\epsilon_0 \left[\left(-1+\left|(r-r_0)\right|^{2}\right) -  2.5/(2 r_0^2) \cdot \left(-1+\left|(r-r_0)\right|^{4}\right)  \right].
\label{eq:potentialCorr}
\end{equation}

	A ready-to-use Perl script for generating a \texttt{table-style} improved inter-atomic potential, with a potential file itself and a description of the method used for computing it is available for download \cite{MendeleyData0}. However, as we realised during our simulations, the newer version of potential did not contribute in any noticeable way to quality of our simulations. Hence, while it may be desirable to use an improved version of potential, in particular when n is different than 2, the version as described in \cite{Koziol0} is sufficiently accurate for reproducing our results.
	
	In simulations of the long-time dynamics of the CSM model, we use the parameter $\epsilon_0$=1 eV and the FCC lattice constant the same as in steel, $a$=3.56 {\AA}, with the mass of atoms assumed as that of Fe. A value  $\epsilon_0$=1.26 eV allows us to reproduce the basic elastic constants and sound velocities of steel 310S, as obtained by using Artur EAM inter-atomic potential \cite{Koziol0}, \cite{Koziol2}.
	
	\section{Analysis of stress dynamics using exact analytical solutions.}
	\label{PaterAnalysis}
	
	Pater's \cite{Pater} exact analytical results reproduce properly the expected dynamics in the limit of continuous medium. In that case, for instance, the wave entering an elastic material exposed to surface Heaviside-type pressure, continuous to penetrate, preserving its original shape. In the case of discrete media, such as the situation of our interests, the Pater's solutions of stress propagation equations are given in terms of Bessel functions. For instance, displacements $u^H_n$, velocities $v^H_n$ and force $f^H_n$, acting on a mass of n-th layer, is given by \ref{eq:Hun}, \ref{eq:Hvn}, and \ref{eq:Hfn}:

	\begin{equation}
		u^H_n = \frac{F_1}{\rm{m}\Omega^2}\cdot\left[J_{2n}(2\theta) + \sum _{k=2n+2,2n+4,...} ^{\infty} (k-2n+1) J_k(2\theta)\right],
		\label{eq:un0}
	\end{equation}
	
	\begin{equation}
		v^H_n = \frac{F_1}{\rm{m}\Omega}\cdot\left[J_{2n-1}(2\theta) + 2\cdot \sum _{k=2n+1,2n+3,...} ^{\infty} J_k(2\theta)\right],
		\label{eq:vn0}
	\end{equation}
	
	\begin{equation}
		f^H_n = -F_1 \cdot \left[ 1-J_0(2\theta) -2\cdot \sum _{k=2,4,...} ^{2n} J_k(2\theta) + J_{2n}(2\theta)\right],
		\label{eq:fn0}
	\end{equation}
	
	where $J_n$ are Bessel functions of the first kind and $\theta = \Omega t$, \rm{m} is the mass of a particle/layer, and $F_1$ is the force exerted on the uppermost layer. The parameter $\Omega$ used there is the fundamental angular frequency of oscillations of two layers, and it is related to the curvature of the potential energy well. The method of determining $\Omega$ has been described by us in detail in \cite{Koziol2}, and in particular in \cite{Koziol0}. These equations are valid for $t<T_N/4$, where $T_N$ is the period of oscillations of, e.g., in a sample's surface pressure $P_{xy}(t)$: A time $t=T_N/4$ is needed for the sound wave to travel in the Y-direction between the upper sample surface and its bottom layer. Hence, that time is proportional to the sample size, i.e., the number of layers N in the Y-direction.
	
	The equations \ref{eq:un0}-\ref{eq:fn0} are derived in \ref{AppendixA} as well, following the work of de Pater. Additionally, we provide there a derivation - for the first time - of similar equations valid when a linear-in-time pressure is applied at the sample surface, when a so-called ramp mode is used in MD simulations.
	
	Some discussion and example use of equations \ref{eq:un0}-\ref{eq:fn0} can be found in 
	\cite{Koziol0}. In this section, we will provide more examples related to the properties of steel 310S with a realistic inter-atomic potential, and compare simulation results between two methods of simulations. Our main aim is, however, a discussion of the propagation of stress in crystals with an ideal harmonic inter-layer potential at very long times, on samples with a large number of crystallographic layers N, and that is done in the next section \ref{DynamicsProperties}. 
	
	\subsection{Computation of Bessel functions.}
	\label{Bessel calculations.}
	
	Performing analytical analysis of MD simulation data with the use of Eqs. \ref{eq:un0}-\ref{eq:fn0} is not as complex as it might look. First, we create a set of data-files (for use with \texttt{Gnuplot}). 
	
	One file contains computed Bessel functions alone: the first column of the data file is the argument $x$ of Bessel functions, and every next column (\texttt{tab}-separated) contains values of functions $J_k(x)$, in increasing order of $k$ (it is not a problem to have a few hundreds of data columns). 
	
	Additional data files contain displacement, velocity, and force computed using Eqs. \ref{eq:un0}-\ref{eq:fn0}, with every next column of the data for every next layer $n$.
	
	In \texttt{Gnuplot}, only up to the 7th layer can be modelled with sufficient accuracy when using \texttt{besjn(n,x)} function available there. \texttt{SageMath}\footnote{\href{https://www.sagemath.org/}{SageMath - Open-Source Mathematical Software System}} also offers comparable accuracy only. For a reliable computation of Bessel functions and quantities dependent on them, we used \texttt{Math::Cephes}\footnote{\href{https://metacpan.org/dist/Math-Cephes/view/lib/Math/Cephes.pod}{Math::Cephes - perl interface to the cephes math library}} \texttt{Perl} libraries. 
	
	According to the documentation, that library provides sufficient accuracy for Bessel function of the order $k$ up to 500, and for their argument, $x$ up to 500.
	In our case, we used successfully $k$ up to 600 and $x$ up to 1000. However, some results break up earlier. The maximal layer number that is modelled properly is around 290. It is easy to notice when calculations with Bessel functions become wrong: a strong, abrupt deviation from a regular dependence on time (i.e., on $x$, the argument of Bessel functions) is observed. 
	
	A ready-to-use Perl script for the generation of data files using Eqs \ref{eq:un0}-\ref{eq:fn0}, with a set of data files created with this description are available for download \cite{MendeleyData1}.
	
	\subsection{Constructing surface pressure curves.}
	
\begin{figure}[ht]
		\centering
		\includegraphics[scale=0.8]{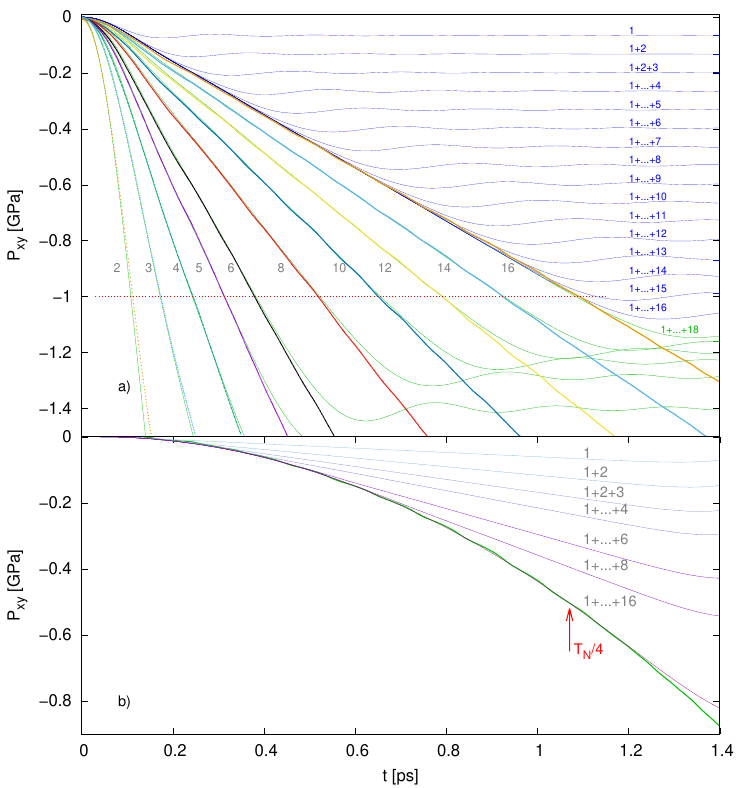}
		\caption{ a) Surface pressure under 1 GPa Heaviside impulse, determined for samples of steel 310S with different numbers of layers, from N=2 to N=16. The tiny green lines in the background are the result of computation using Eq. \ref{eq:sumfn}. 
			b) The green line shows $P_{xy}(t)$ simulation results of surface pressure under ramp rate h=1000 MPa/ps, found for the same sample of steel with N=16 layers.
	At t=$T_N$/4 the front of the pressure wave reaches the opposite side of the sample (shown by the red arrow). The labels indicate the range of layers over which summation is performed. 
		}
		\label{fig:PressureScaling310S00}
	\end{figure}

	Since pressure at the surface $P_{xy}$ is a volume average of the internal virial stress, we can write:

	\begin{equation}
		P_{xy}(t) = \sum_{m=1,2,...} ^{n} f_m(t),
		\label{eq:sumfn}
	\end{equation}
	
	where the maximum summation range $n$ should not exceed the number of layers N, and it ought to be carried for times $t < T_N/4$, i.e., before the front of pressure reaches the opposite side. Computational results of $P_{xy}(t)$ using this method are shown in Fig. \ref{fig:PressureScaling310S00} a), for the Heaviside-type of force applied, for samples of steel 310S with different number of layers, from N=2 to N=16. A comparison with MD simulations is included there. 
	
	The period of pressure oscillations in this case is 4.28 ps for the sample with N=16. At $t=T_N/4$ the front of the pressure wave reaches the opposite side of the sample (shown by a red horizontal broken line at a value of -1 GPa), penetrating the whole sample volume. We use successive approximations of $P_{xy}(t)$ based on Eq. \ref{eq:sumfn}, by summing a finite number of contributions from forces $f_{\rm{m}}(t)$, where \rm{m} denotes the number of layer counted from the surface of the sample, where the force is applied. The grey labels indicate the number of layers in a sample, N.
	The blue labels indicate the range of layers over which summation is performed. 
	The parameter $\Omega$ used here is 14.45/ps.
	In very short times, less than the time needed for the transverse sound wave with speed $c_T$ to pass the distance between two layers, which is around t=2.06{\AA}/c$_T$ {\AA}/ps = 0.068 ps, the only first contribution, $f_1$ of Eq. \ref{eq:sumfn}, is sufficient for approximation of the surface pressure. There, c$_T$=30.3 {\AA}/ps is the transverse speed of sound, determined separately. 
	
	A similar method is used in Fig. \ref{fig:PressureScaling310S00} b) for drawing $P_{xy}(t)$ in the case of a linear-in-time ramp of the surface pressure, $P^0_{xy}(t)=h\cdot t$, with ramp rate h=1000 MPa/ps. In this case, we use Eq. \ref{eq:Lfn4} on forces $f^L_m$ acting on layers. As we see, 	$P_{xy}(t)$ dependence strongly resembles a parabolic one. In fact, we are able to show that in the limit of N $\gg$ 1, $P_{xy}(t)$ curves observed in linear-in-time ramp pressure can be constructed as a sum of a linear-in-time and parabolic components.
	
	The diagrams as these in Fig. \ref{fig:PressureScaling310S00}, as well as the dependencies of velocity and displacement or virial stress of layers over time all require only two fitting parameters. The one is related to the amplitude (of, for instance, pressure or velocity), and the other parameter determines the timescale. The first one is related to coefficients in front of Pater's equations \ref{eq:un0}-\ref{eq:fn0}, for the Heaviside boundary condition, and in front of equations \ref{eq:Lun3}, \ref{eq:Lvn}, and \ref{eq:Lfn4}, for the linear-in-time change of the surface force. The second parameter is related to $\Omega$.
	
	\subsection{Velocity of layers.}
	\label{Velocity of layers}
	
\begin{figure}[ht]
		\centering
		\includegraphics[scale=0.8]{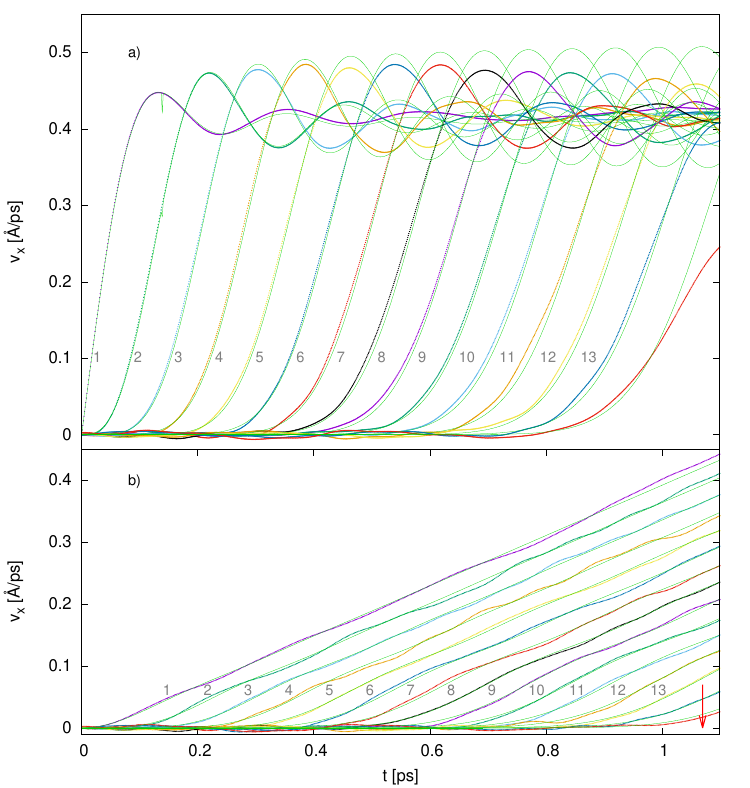}
		\caption{Velocity for the same sample as in Fig. \ref{fig:PressureScaling310S00}. a) is for the Heaviside type of pressure function, and b) is for ramp rate h=1000 MPa/ps. The parameter $\Omega$ used here for drawing computed curves is 14.45/ps, the same as in Fig. \ref{fig:PressureScaling310S00}. Tiny green lines are drawn using Eq. \ref{eq:Lvn}. Labels indicate the number of layer.
		}
		\label{fig:Bessel02}
	\end{figure}

	The velocity of layers for a linear-in-time ramp rate of h=1000 MPa/ps is shown in Fig. \ref{fig:Bessel02}. This is the same sample and the same simulation conditions as used in Fig. \ref{fig:PressureScaling310S00}. The analytical results have been computed with Eqs. \ref{eq:vn0} and \ref{eq:Lvn}, for the Heaviside and the linear-ramp case, respectively.
	An oscillating contribution to velocity is more pronounced in MD simulation results than that computed with Bessel functions. 
	In the case of ramp pressure, as a best fitting parameter value determining the velocity scale, we used 0.0282 {\AA}/ps, while the value computed from Eq. \ref{eq:Lvn}, $h/(\rm{m}\Omega^2)$ is 0.0287 {\AA}/ps (surface area per atom is 5.49 {\AA}$^2$ and average mass of atoms is 9.211$\cdot$10$^{-26}$kg). Hence we obtain a good agreement between calculations based on analytical formulas and the results of MD simulations.
	
	\subsection{Virial stress.}
	\label{Virial stress}
	
\begin{figure}[ht]
		\centering
		\includegraphics[scale=0.8]{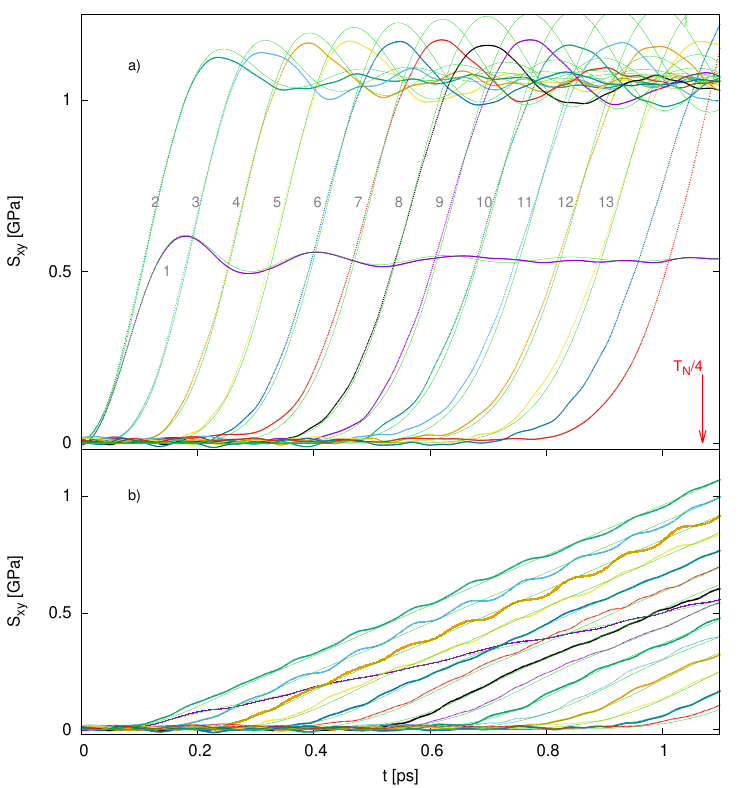}
		\caption{Virial stress for layers, determined in the same measurements as these in Figures \ref{fig:PressureScaling310S00}, and \ref{fig:Bessel02}. a) is for Heaviside-type pressure and b) when ramp force is applied, with h=1000 MPa/ps. Labels indicate the number of layer. $T_N$/4 is marked by red arrow.
		}
		\label{fig:Bessel03}
	\end{figure}

	In \texttt{lammps}, the entire force acting on layers is reported, while Pater's equations provide force acting on one side of the layer, only. Therefore,
	it is argued \cite{Koziol0}, to compare the results of \texttt{lammps} simulations and Pater's, we need to take a sum of two results from Pater's equations for neighboring planes. There is an exception to that rule. In the case of the uppermost layer, the force in \texttt{lammps}' simulations acts on the layer from one side only (since there is no layer above it). The same is in the case of the lowermost layer. Therefore, in these cases, the Pater's result for one layer only ought to be used. This explains why the virial stress on the layer labeled as 1 (the uppermost) in figure \ref{fig:Bessel03} is about 2 times smaller than that for the remaining layers.
	
	\subsection{Displacement of layers.}
	\label{Displacement of layers}
	
\begin{figure}[ht]
		\centering
		\includegraphics[scale=0.8]{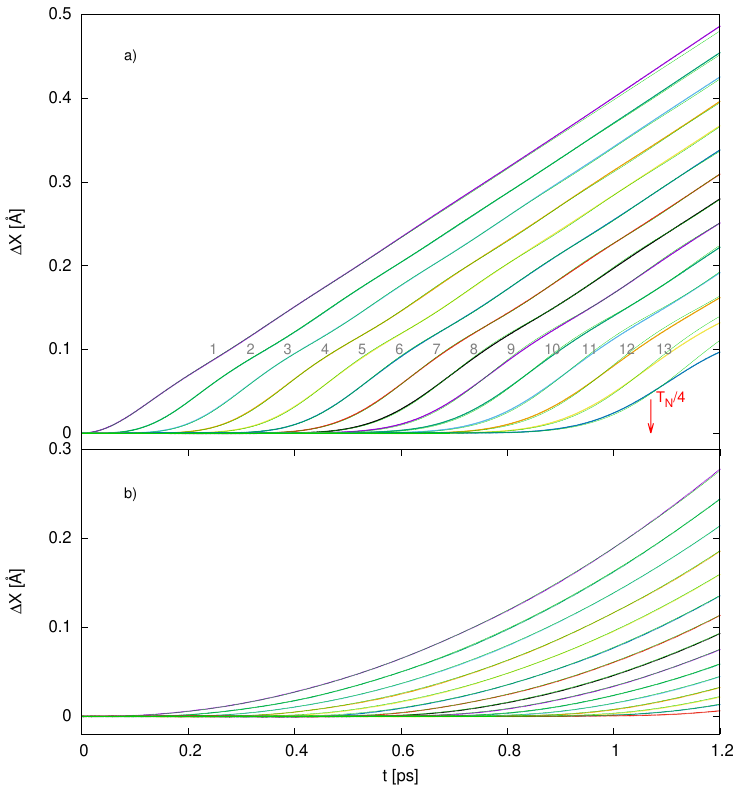}
		\caption{Displacement of layers (with their number shown by labels) for Heaviside-type pressure applied, in a), and for ramp rate of h=1000 MPa/ps in b), determined for a sample of steel 310S with N=16 layers. $\Omega$=14.4/ps was used, and the displacement scaling factor in b) is 0.00196 {\AA}, which is in very good agreement with $h/m\Omega^3 = 0.001996${\AA} computed by using Eq. \ref{eq:Lun3}.
		}
		\label{fig:Bessel04}
	\end{figure}

	While velocity and virial stress for layers, in the case of linear ramp pressure imposed, may be seen as linearly dependent on time (when neglecting superimposed small oscillations), the displacement of layers follows very closely a parabolic dependence on time, as seen in Fig. \ref{fig:Bessel04} b), while in the case of Heaviside pressure it is a linear function of time, as shown in a).
	
	For determining curves like these from MD simulations, one needs to compute the average position of atoms, as reported by \texttt{lammps}, by taking into account their image index. Otherwise, when atoms move outside of the simulation \texttt{box}, a jump in average values of their positions would be found.
	
	\subsection{MD temperature.}
\label{MDTemperature}

Equations \ref{eq:un0}-\ref{eq:fn0} allow us to determine much more than displacements, velocities, and forces. In \texttt{lammps}, temperature is computed from the kinetic energy $E_k$ of the system, and in the case of ours, when particles possess 3 degrees of freedom of the movement, it is given by the relation: $3/2 \cdot k_B \cdot T \cdot N_{atoms} = E_k$, where $N_{atoms}$ is the total number of atoms. On the other hand, the kinetic energy is related to the velocity of particles. We will show that, contrary to the common misconception, in simulations like these we are dealing with, the temperature as reported by \texttt{lammps}, is not a thermodynamic quantity. It is related to the velocity of layers. 

\begin{figure}[ht]
	\centering
	\includegraphics[scale=1.0]{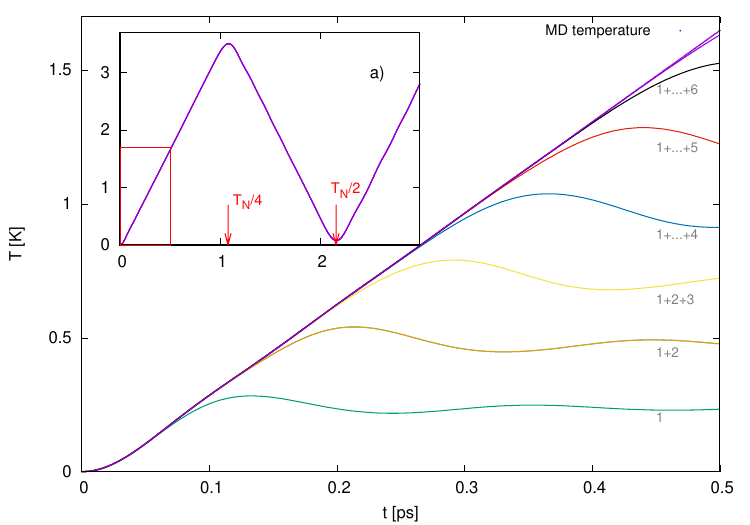}
	\caption{Temperature, as reported by \texttt{lammps}, when a Heaviside-type simulation
		is carried out at a pressure of 1000 MPa. No temperature stabilization is carried out, while the initial temperature of the sample was 0K.
		The labels show the summation range, over which layers the kinetic energy is computed, by using Eq. \ref{eq:T}.
		Inset a) shows the same T(t) dependence in an expanded range; the red rectangle is the region shown in the main figure. Red arrows indicate 1/4 and 1/2 of the period $T_N$.
	}
	\label{fig:Bessel_T00}
\end{figure}

To show that, we did as follows. We cooled down to $T=0$ the same sample as used for obtaining the data shown in figures \ref{fig:PressureScaling310S00}-\ref{fig:Bessel04} (there, the sample temperature was stabilised at 50 K during simulations). This time we do not stabilise temperature. Computation is done at the Heaviside-type of simulations with the value of pressure applied of 1000 MPa.
Let us compute the kinetic energy of layers by using Eq. \ref{eq:un0}. It will be given as a sum of the kinetic energies of all $N$ layers: $E_k= \sum_{i=1}^{N} E_{k,i}$, where $E_{k,i}$ is the kinetic energy of the i-th layer. Obviously, $E_{k,i}= \rm{m} \cdot v_i^2/2 \cdot N_{layer}$, where \rm{m} is an average mass of one atom, $N_{\rm{layer}}$ is the number of atoms in one layer, and $v_i$ is given by \ref{eq:vn0}. Results of Fig. \ref{fig:Bessel_T00} show that temperature computed from the kinetic energy may be written as:

\begin{equation}
T  = \frac{2E_k}{3 k_B N_{atoms}} =
 \frac{(P^0_{xy} \cdot S)^2}{3k_B \rm{m}\Omega^2 N}
 \sum_{i=1}^{N}
 \left[J_{2i-1}(2\theta) + 2\cdot \sum _{k=2i+1,2i+3,...} ^{\infty} J_k(2\theta)\right]^2,
\label{eq:T}
\end{equation}

where we used Eq. \ref{eq:vn0}. With the surface area per atom, $S=a^2\sqrt{3}/4$=5.49$\cdot$10$^{-20} m^2$, the applied pressure $P^0_{xy}$=1000 MPa, the number of layers N=16, and $\Omega$=14.45/ps, the average mass of the atom m=9.211$\cdot
 10^{-26}$ kg, the prefactor in \ref{eq:T} has a value of 0.2373 [K]. The value of the scaling factor used by us, determining the magnitude of temperature in Fig. \ref{fig:Bessel_T00}, is 0.23726 [K]. 
 
 An interesting property of $T(t)$ dependence in the case of the Heaviside method is that the period of oscillations in temperature is twice shorter than  $T_N$, as is seen in Fig. \ref{fig:Bessel_T00} a).
	
We arrive  at a few useful conclusions. In simulations like those, the temperature reported by \texttt{lammps} must not be considered as a thermodynamic temperature. Instead, the reported temperature may provide valuable insight into the kinetics of ongoing dynamic processes. Moreover, stabilising temperature will introduce changes in the kinetic energy of layers, disturbing the quality of the simulation results, and therefore it ought to be avoided. An equally important is the next conclusion: we are able to relate the entire kinetic energy of the system with the computed velocities of layers in the x-direction only. That means no other components of velocity provide any significant contribution to the kinetic energy; in particular, there is no noticeable thermodynamic contribution caused by (random) motion of atoms. We performed a separate analysis of averaged velocities in the y- and z-direction as well, finding that these are around 2 orders of magnitude smaller than velocity in the x-direction. It is worth to mentioning that similar observations hold in cases when dislocations are present (below the pressure values causing their movement). 

Hence, the fundamental assumption of the CSM model, which is treating atoms in crystallographic layers as sets of entities moving together in one direction is confirmed by the above observations.

	\section{Large N and long time dynamics. Properties.}
	\label{DynamicsProperties}
	
	\subsection{Simulation details and general characterization of dynamics.}
	\label{Simulation details}
	
	The MD simulation results, as shown for steel, such as the dependence on time of virial stress or velocity of layers, indicate that there are non-linearities present in the interaction of layers. These effects are noticeable in values of pressure that we have considered so far very low. As a consequence of these effects, the results of simulations suggest the presence of seemingly chaotic dynamical processes and a possible instability of crystal structure when simulations were performed at long times and/or on samples with a very large number of layers. For a proper study of these effects it is therefore reasonable to use a harmonic interatomic (interlayer) potential and samples with uniform chemical composition (i.e., containing one atom type).

\begin{figure}[ht]
		\centering
		\includegraphics[scale=1.0]{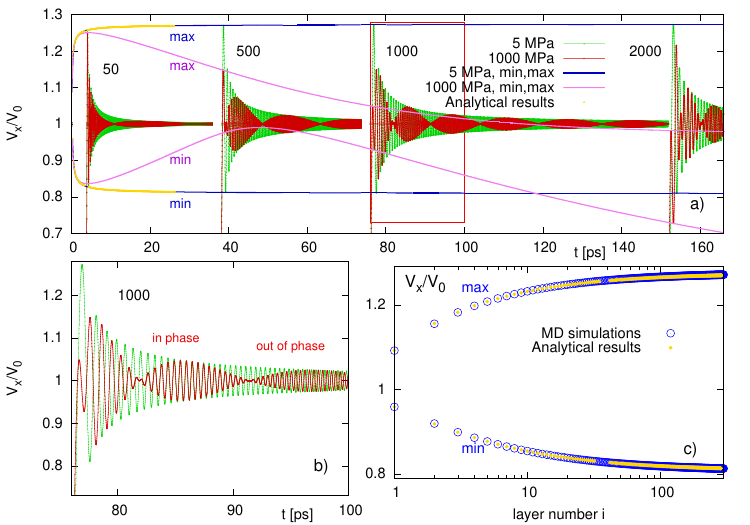}
		\caption{
			Velocity of layers is normalised by $v_0$, for a few layers (50, 500, 1000, and 2000), as shown by labels, for simulations performed under pressures of 5 and 1000 MPa (green and red curves, respectively). The time step used there is 0.5 fs in both cases. The envelope functions (blue and magenta curves, the blue one for 5 MPa, and magenta for 1000 MPa) show the limits of the amplitude change in oscillations. The one marked as $max$ shows the position of the first maximum, and $min$ is the position of the first minimum in $v(t)/v_0$. The envelope functions were determined for all layers in the sample (N=15000 in this case).
			The red rectangle in a) marks the region expanded in b), for the layer number 1000. 
			In a) the golden points covering blue lines are computed analytically by using Bessel functions (Eq. \ref{eq:Hvn}), hence this is a demonstration that positions of the first maxima and minima in velocity drawn as a function of time coincide with these from MD simulations. In c) we show the same data drawn as a function of layer number (up to 290 layers), with log scale on horizontal axis for a better visualisation of accuracy.
		}\label{fig:LONG_H_VxCh0X3}
	\end{figure}

	Figure \ref{fig:LONG_H_VxCh0X3} illustrates results of simulations performed at 5 MPa and at 1000 MPa. We clearly see the dependence of results on the value of pressure, an effect that is not present in the linear model of Pater. 

	The envelope functions in Fig. \ref{fig:LONG_H_VxCh0X3} (blue and magenta curves) show the limits of the amplitude change in oscillations. The one marked as $max$ shows the position of the first maximum, and $min$ is the position of the first minimum in $v(t)/v_0$. When large pressure is applied, the envelope function and the resulting oscillation patterns in $v(t)/v_0$ may assume forms hard to interpret.
The red rectangle in a) marks the region that is expanded in the lower part b), for layer number 1000. A strong beat in oscillations for 1000 MPa are observed, with regions in phase and out of phase, repeating sequentially. 

As an illustration of the sensitivity of the studied dynamic processes to pressure values, let us notice the following: the slope of the $P_{xy}(t)$ curve (normalised by the applied pressure, which is different by 200 times) changes by less than $10^{-5}$ when pressure is changed between 5 and 1000 MPa.

It is also worthwhile imagining how small the displacements are during these simulations. When pressure of 5 MPa is used, the maximal displacement between the nearest layers is around $1.7\cdot 10^{-5}$ {\AA} and the value of velocity $v_0$ in Fig. \ref{fig:LONG_H_VxCh0X3} is $2.25 \cdot 10^{-3}$ {\AA}/ps. 

	Our first idea was that these unexpected and undesirable effects must be due to deviations of the interatomic potential (strictly of the interlayer potential) from the harmonicity. For that reason, we created an improved potential that compensates for small $x^4$ contribution to $x^2$ dependence. The results obtained with an improved potential were, however, identical to these with the original potential as used
	in \cite{Koziol0}.
	
	Due to other tests, we found out that the frequency beats effect is caused by the movement of layers in Y- and Z-directions. While that movement  is very small, at long times, and for samples with large N, the effect is pronounced. It can be suppressed by using a \texttt{fix} in \texttt{lammps} scripts disallowing the movement of atoms in directions other than X. That does not influence the quality of our simulation results since we are aiming in comparison of simulations with an analytical 1-dimensional model. In any case, when simulations are carried out at very low pressure values; their outcome does not depend whether a restriction is imposed or not on the movement of atoms in y- and z-directions.
	We have performed an extensive analysis of the effect of beats in frequency; however, we consider providing a detailed description of it outside of the scope of this article. 
	
	Another critical parameter in any simulations like these is the time-step \texttt{ts} used. On one hand, it is preferred to have a long time step for faster computation. However, too long time step will eventually lead to improper integration of equations of motion by \texttt{lammps}. We observed that in our case, when \texttt{ts} is shorter than around 2 fs, no undesirable effects are present. The results presented here have been obtained for \texttt{ts} of 0.5 fs, and additionally checked for agreement with those obtained on a shorter sample by using \texttt{ts}=0.2 fs.

	The present results have been obtained on samples with 15000 layers (3 $\mu m$ in size), by using a very small applied pressure of 5 MPa, in order to minimise any undesired, complex dynamic effects. 
	
	Simulations on samples of that size take a long time and produce a large amount of data for analysis (over 3 TB altogether in our case), and the post-processing analysis of the data is also time-inefficient. Therefore, for initial testing, it was necessary to resort to the use of samples of size, e.g., with N=3000 layers, and only the final results are shown for a sample with N=15000 layers.
	
\begin{figure}[ht]
		\centering
		\includegraphics[scale=1.0]{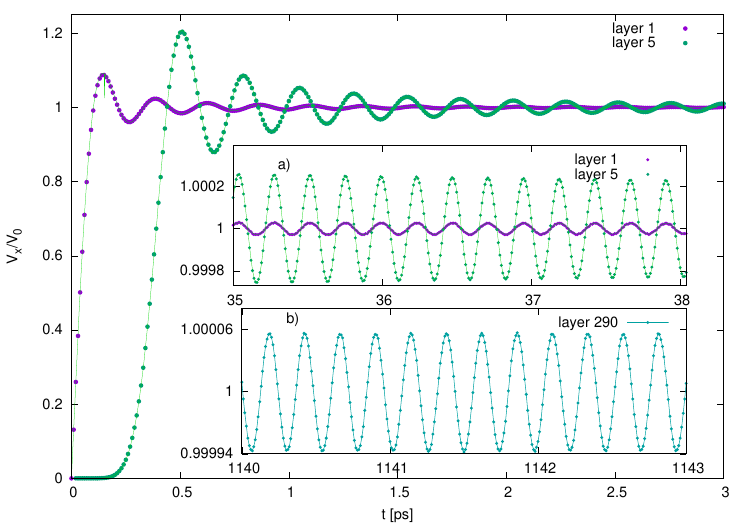}
		\caption{
			Comparison of MD simulation for layers 1 and 5 with the computational results using Bessel functions, in short times.
			In a) we show the same kind of data near the largest time we are able to compute Bessel functions (shown by tiny green line over the data points) for layers 1 and 5, which is 38.4 ps. 
		}\label{fig:LONG_H_VxCh00}
	\end{figure}

Figure \ref{fig:LONG_H_VxCh00} shows an example comparison of MD simulation results and computational curves obtained by using Bessel functions. 
	For layer 290 we are able to compute Bessel functions, only up to 23 ps. The MD simulation results as shown in b) for layer 290 are in the vicinity of the time when the stress wave reaches the bottom of the sample, which is at $t=T_N/4 \approx$1140.8 ps.

	\subsection{Stretched-exponential scaling of oscillation period.}
	\label{Stretched}

\begin{figure}[ht]
		\centering
		\includegraphics[scale=1.0]{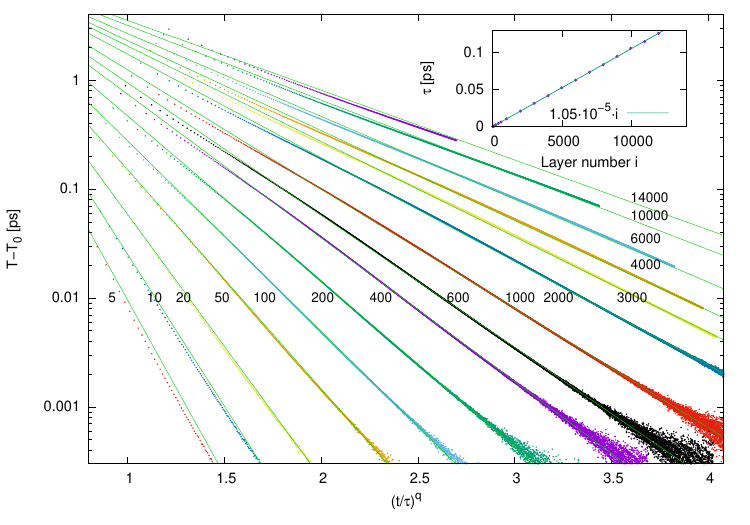}
		\caption{
			Period of oscillations $T$ as a function of the power of time. 
			Labels indicate the layer number. 
			The data were drawn with fixed values of $A=15.5$ ps and $q=0.203$, while $\tau$ was searched in least-squares fitting procedure. It was found (as the inset of figure shows) that $\tau$ is a linear function of layer number $i$: $\tau$ = $1.05 \cdot 10^{-5}\cdot i$ [ps].
		}\label{fig:LONG_H_VxSound05}
	\end{figure}

	It is evident that while the amplitude of oscillations in $v(t)/v_0$ decreases with time,
	figures \ref{fig:LONG_H_VxCh0X3}, \ref{fig:LONG_H_VxCh00} show us 
	that the period decreases strongly with time as well. Let us define that period $T$ as 
	a time difference between two successive moments when $v(t)/v_0-1$ crosses upward the value of 0. The results of such analysis are shown in Fig. \ref{fig:LONG_H_VxSound05}. 
	
	$T$ changes mostly initially, when a pressure wave reaches a given layer, and asymptotically, at long times, that period tends to reach a constant value $T_0$. With $T_0$=0.239 ps, $\omega_0$ defined as $2\pi/T_0$ has a value of 26.2895 /ps, which is surprisingly close to $2\cdot\Omega$ for our sample ($\Omega$ has been determined separately to have a value of 13.144/ps). Therefore, we assume that, indeed, at long times, the period of oscillations becomes related to the argument of Bessel functions, which is $2\cdot\Omega t$. Some sets of data points (for i>2000) end up well before the lowest values of $T$ are reached. This is due to a limited range of time span of simulations performed, up to 1200 ps.
	
	This figure shows that the period of oscillations $T(t)$ tends to reach $T_0$ at long times and $T(t)$ can be well described by a stretched-exponential function, $T(t)-T_0=A\cdot exp(-(t/\tau)^q)$: when it is drawn with the log scale on the vertical axis as a function of $(t/\tau)^q$, at the proper choice of the power parameter $q$, a straight line is obtained. It is well known that finding the parameters of the stretched-exponential function is usually ambiguous to some degree. Our attempts to perform least-squares fitting gave a significant spread in values of $A$ and $\tau$ while a rather well-defined value of $q$ was obtained. The data shown in this figure is drawn with fixed values of $A=15.5$ ps and $q=0.203$, while $\tau$ was searched in fitting procedure. It was found (as the inset of figure shows) that $\tau$ is a linear function of layer number $i$. Let us summarise these observations with the following equation:
	
	\begin{equation}
		T(t)-T_0=A\cdot exp(-(t/\tau)^q),~~~\tau = 1.05 \cdot  10^{-5}\cdot i ~[\rm{ps}].
		\label{eq:kohl}
	\end{equation}
	
	It remains unclear to us how to interpret the physical and/or mathematical values of parameters $A, q$ and $\tau$. The significance of the stretched-exponential time relation (known also as Kohlrausch law) is that it has never been used, to the best of our knowledge, to a description of processes in a perfectly ordered medium. It has been used instead to characterise a broad class of phenomena that are governed by some kind of disorder and statistical probability. It is most often used to describe relaxation effects in dielectric materials 
	\cite{Milovanov} or luminescence decays \cite{Berberan}, \cite{Bodunov}, 
	or magnetic susceptibility in spin-glasses, and it is used often in studies of phenomena in nature, economy, and sociology \cite{Laherrere}.
	
	The only common property between those where the dynamics of materials/phenomena is analyzed by the Kohlraush law and our observations is that there are some collective interactions involved in all these cases. A few only authors attempt to analyse the origins of Kohraush law, by going in that direction \cite{Kliem}. 
	
	A puzzling property is the linear relation between the parameter $\tau$ and the number of layer $i$ in Eq. \ref{eq:kohl}.

	\subsection{Speed of sound and universal scaling of layer dynamics.}
	\label{Sound}
	
	An intriguing question is how to find out the speed of sound for this system. A simple idea is to measure the distance between layers and to find out the time difference between the arrival of the pressure wave to these layers. For that, we must assume a certain limiting value of displacement, or virial stress, or velocity to be reached as an indicator of the arrival of a pressure wave. In \cite{Koziol0} we assumed that 0.1 of virial stress must be reached. Here, we choose to investigate the problem more deeply. 
	
	\begin{figure}[ht]
		\centering
		\includegraphics[scale=1.0]{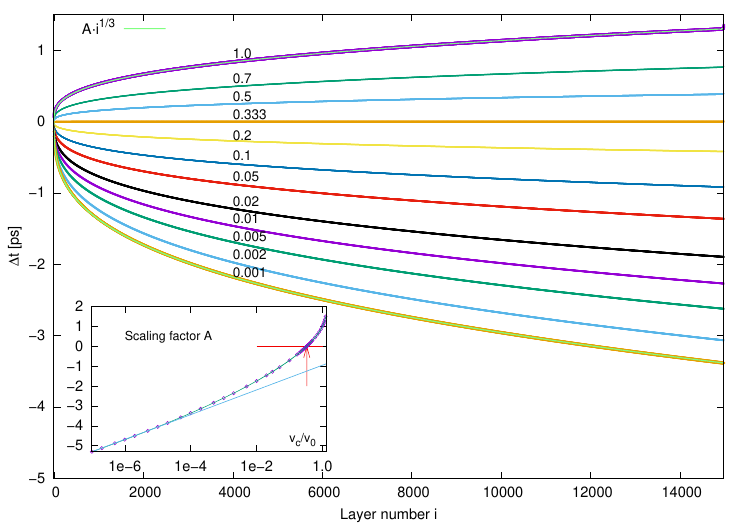}
		\caption{
			The time difference $\Delta t$ for layers $i$ from 1 to 15000 between the value of $v_c/v_0$ as shown by labels (from $v_c/v_0$=1.00 for the top curve to $v_c/v_0$=0.001 for the lowest curve), and the time when $v_c/v_0$=0.333 is observed for any given layer. The value of $v_c/v_0$=0.333 is chosen as the reference, because we observe that the sound speed determined by using that value is found to be a function independent of time.
		}\label{fig:LONG_H_VxSound00B}
	\end{figure}

	It ought to be emphasised that analysis of the kind we are presenting here (but also in other parts of this work) was possible only due to possessing a very well-defined set of data and by using self-developed scripts designed for semi-automatic data analysis. 
	
	There are a few steps in this analysis.
	
	Our idea was to find out what the time is when the velocity of a layer, $v(t)$, normalised by $v_0$, that is $v(t)/v_0$, crosses for the first time a given value $v_c/v_0$. We wanted to have the data for all layers (15000 in this case). After that, we draw the dependence of the obtained time as a function of layer number.
	
	That, effectively, gives us a set of curves on the dependence of sound propagation distance on the layer number (i.e., on the distance). These curves were almost linear. But the linearity between time and distance was not perfect. 
	
	Surprisingly, we found out (and that is not enforced by any mathematical rule known so far to us) that in the case when $v_c/v_0=1/3$ is chosen as the limiting value, then the speed of sound remains constant.
	
	For the above reason, we assumed the curve of change of time when $v_c/v_0=1/3$ as the reference one.
	
	Any other curves, for $v_c/v_0$ not equal to 1/3, are considered as a departure from it.
	When the criterion $v_c/v_0=1/3$ is used, the difference of time arrival of the pressure wave to layer $i$ in comparison to that for layer $i=1$ is given as: $\Delta t = 0.07606732 \cdot (i-1)$ [ps].

	In that way, the results in Fig.  \ref{fig:LONG_H_VxSound00B} show us a time difference between the time moments when the velocity of a layer achieves a certain value and the time when its value becomes 1/3 of $v_c/v_0$. For instance, the outermost curve labelled as "1.0" shows the time difference between $v_c/v_0$ reaching the value of 1, and the time when it reaches the value of 1/3.
	
	Additionally, we found out that, with high accuracy, all the curves shown are described well by a simple function of the layer number $i$: 

\begin{equation}
	\Delta t(i) = A\cdot i^{1/3}, 
\label{eq:Dt}
\end{equation}

	with the coefficient $A$ depending on $v_c/v_0$. 
	
	The value of $A$ is 0.052756 ps for the uppermost curve (when $v_c/v_0$=1), and -0.1372 ps when $v_c/v_0$=0.001. The function expressed with \ref{eq:Dt} is drawn with tiny green lines on top of the data points for both these outermost curves. 
	
	The inset in figure \ref{fig:LONG_H_VxSound00B} shows how the scaling coefficient $A$ changes with 
	$v_c/v_0$. There, $A$ is normalised to 1, corresponding to a value of $A$=0.0527559 ps for the uppermost curve in Fig. \ref{fig:LONG_H_VxSound00B}. The $A(v_c/v_0)$ curve changes sign at a value of $v_c/v_0$ which is very close to 1/3 (more accurately, our best fit gives us 0.332984). At $v_c/v_0 < 10^{-4}$ the dependence is well approximated by a logarithmic one, as illustrated by the straight line. Notice that the inset figure is drawn for values as low as $10^{-7}$ of $v_c/v_0$, hence, the velocity there has been determined with an accuracy exceeding $10^{-10}$ {\AA}/ps.
	
\begin{figure}[ht]
		\centering
		\includegraphics[scale=1.0]{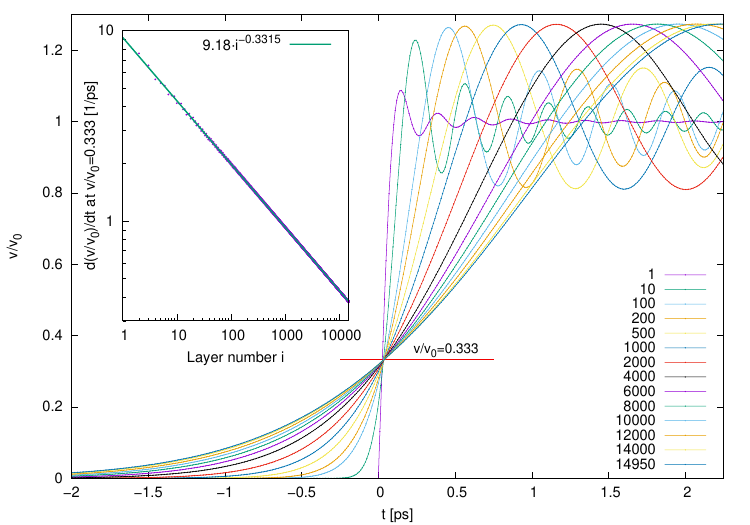}
		\caption{
			Change of slope of velocity curves for different layers, as shown by the legend). 
			The inset figure shows that the derivative of the slope of $dv(t)/dt$ at $v/v_0=0.333$ changes with the layer number $i$ as $i^{-0.3315}$.
	}\label{fig:LONG_H_Vx_SLOPE}
	\end{figure}

	As a consequence of the observation that $\Delta t = A\cdot i^{1/3}$ in Fig. \ref{fig:LONG_H_VxSound00B}, we realised that the slope of curves of $v(t)/v_0$ at around $v_c/v_0=1/3$ may undergo a similar scaling rule with layer number. Indeed, results of Fig. \ref{fig:LONG_H_Vx_SLOPE} show that the slopes of curves change as $i^{-0.3315}$:
	
\begin{equation}
 \frac{dv(t)}{v_0 dt} \bigg| _{v/v_0=0.333} = B\cdot i^{-0.3315}, 
\label{eq:Dv}
\end{equation}
	
	with $B$=9.18/ps. The value of parameter $B$ must be related to $\Omega$, only. 

Virial stress curves have practically an identical dependence on time as velocity curves, when $t<T_N/4$ and $i$ is not very small, and therefore properties of the virial pressure profiles can be discussed in the same way.

\begin{figure}[ht]
		\centering
		\includegraphics[scale=1.0]{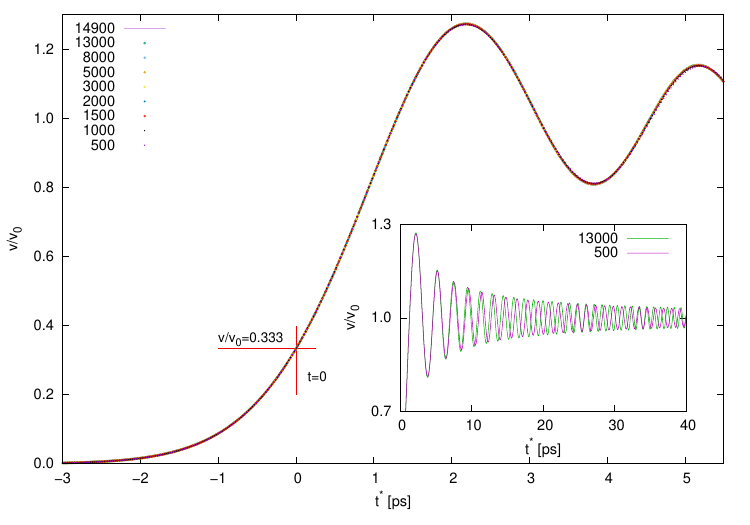}
		\caption{
			Scaling of entire $v(t)/v_0$ curves with time, for a few layers, as shown in the legend. We choose as $t$=0 the point at $v/v_0=0.333$, where curves in Fig. \ref{fig:LONG_H_Vx_SLOPE} intersect. The time transformed as described in the text is named here as $t^*$. The inset figure shows a view of layers 500 and 13000 with an expanded $t^*$ axis.
		}\label{fig:LONG_H_Vx_SLOPE2}
	\end{figure}

As a result of observations explained with the help of figures \ref{fig:LONG_H_VxSound00B} and \ref{fig:LONG_H_Vx_SLOPE}, and expressed by Eqs. \ref{eq:Dt} and \ref{eq:Dv}, 
we conclude that a rescaling of entire $v(t)/v_0$ curves may be possible. As figure \ref{fig:LONG_H_Vx_SLOPE2} shows, this is indeed the case. After changing the time origin and its scale in a proper manner, we find out that all the curves $v(t)/v_0$ can be merged into one.

In the first step, we choose as $t=0$ the point at $v/v_0=0.333$, where curves in Fig. \ref{fig:LONG_H_Vx_SLOPE} intersect. At that point, time has a value of around 0.037 ps. We shift all the curves left for that value. Next, we rescale time for the displayed curves in such a way that they coincide with the data for layer number 14900. This is done by using a function displayed on the inset of Fig. \ref{fig:LONG_H_Vx_SLOPE} and given by Eq. \ref{eq:Dv}, that is by multiplying the time axis for every layer $i$ by the factor $(14900/i)^{0.3315}$. Hence, the timescale for a layer $i$=14900 does not change, while the new, transformed time values for $i\ne$14900 are strongly expanded. That is, for instance, for a layer $i$=500, the time axis is expanded $(14900/500)^{0.3315}\approxeq 3.053$ times. Let us name the new, transformed time as $t^{*}$.

The inset of Fig. \ref{fig:LONG_H_Vx_SLOPE2}, with an expanded view of $v(t^*)/v_0$ for layers 500 and 13000, shows how a departure from this scaling evolves with transformed time $t^*$.  The positions of maxima in $v(t^*)/v_0$ change slowly with $t^*$ and $i$. The amplitude of oscillations at large values of $t^*$ does not, however, depend on $i$, in the first approximation. We suspect that the scaling relation might be a manifestation of an inherent mathematical property of the dynamics of the system, perhaps asymptotically valid exactly in the limit of large time and large N.

	\subsection{Amplitude of oscillations in v(t).}
\label{Positions of maxima}

\begin{figure}[ht]
	\centering
	\includegraphics[scale=1.0]{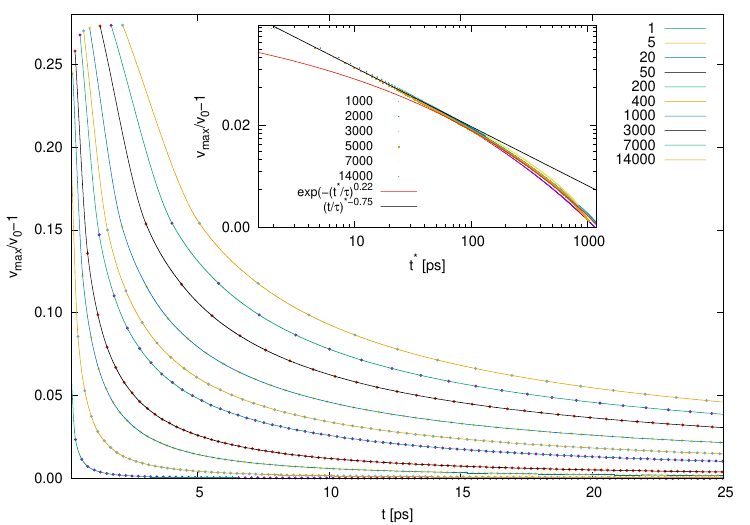}
	\caption{
		Changes of the amplitude of oscillations in $v(t)/v_0$ for a few layers, as described by the legend.
		The inset figure shows how all the data points merge together to form an (almost) single curve after the time is transformed to $t^*$, in the same way as it was in Fig. \ref{fig:LONG_H_Vx_SLOPE2}.
		Two continuous lines in the inset figure show approximations of the data by a power law, and a stretched-exponential function of the transformed time $t^*$.  
	}\label{fig:LONG_H_VxSoundMinMax00A}
\end{figure}

Changes in the amplitude of oscillations in $v(t)/v_0$ may be discussed by analysis of changes with time of the maximal values of $v(t)$, as shown in Fig. \ref{fig:LONG_H_VxSoundMinMax00A}, for a few layers, as indicated in the legend. The data are drawn as points, with continuous curves for eye guidance. 

For a better comparison of the dependences for various layers, the curves have been shifted towards the left side of the graph in the same way as we did for curves in Fig. \ref{fig:LONG_H_Vx_SLOPE2}, however, in this case we do not transform the time variable. 

For analysis of the time dependence of the amplitude, we, however, merge data points from different layers into a single set of data by using the same time transformation as used in Fig. \ref{fig:LONG_H_Vx_SLOPE2}. The results are shown in the inset figure. We find that a power law describes the data better in short times, while a stretched-exponential function is more suitable at the side of long times.

	\section{Summary and conclusions.}
	\label{Summary and conclusions.}
	
	It has been demonstrated that the CSM model, originating from the works of Schr{ö}dinger \cite{Erwin}, \cite{Muhlich} (1914), and de Pater \cite{Pater} (1974), is suitable as an analytical description of the dynamics of layers in orientated FCC crystals. 
	
	An analytical extension of that model has been derived for the case of linear-in-time ramp pressure applied to the sample surface. Exact solutions of the model provide a description of the displacement, forces, and velocities of crystallographic layers in terms of Bessel functions. 
	
	Examples have been provided of using the model in the description of properties of steel 310S, with realistic inter-atomic potentials applied, in cases of the Heaviside-type and linear-in-time ramp pressure. Not shown simulation results on several other medium-entropy alloys studied by us are scalable according to this model as well when the proper adjustment of material parameters is done. These are CoNiCr (with EAM and MEAM inter-atomic
	potentials), and CoNiV (with MEAM potential). 
	
	In order to test the validity of the CSM model, its dynamical stability for large sizes of samples (15000 layers, $3 \mu m$ in size), and for a long time, simulations were performed by using a harmonic interlayer potential. 

	The dynamics of the CSM model of perfectly ordered matter is found to be describable by stretched-exponential time functions, with the dynamics characterised by scaling properties in time and in the size of crystal structure. 
	
	We believe that the model provides a new field for mathematical research. Some of the observed dependencies may have an explanation by using advanced mathematical analysis of de Pater's solutions. 
	
	The model is particularly helpful with understanding dynamical processes in MD simulations of crystals exposed to stress. However, it has potentially broad applications in other fields of material physics, far beyond the realm of classical mechanics.

	\appendix
	\section{Mathematical model: a chain of springs and masses.}
	\label{AppendixA}
	
	\subsection{Introduction.}
	\label{AppendixIntro}
	
	Quantities that can be accurately determined in MD simulations are, in our case, average values for all atoms within the planes perpendicular to [111] direction (Y-direction). These are, for instance, $\langle S_{xy,n}\rangle $, $\langle X_n\rangle $, $\langle V_{x,n}\rangle $ (xy-component of virial stress, displacement and velocity of atoms in X-direction, respectively. $\langle \rangle $ means an average over all atoms within the plane, while index $n$ enumerates planes, starting with $n=1$ for the surface rigid plane, where force is applied, and with $n=N$ for the plane  with the lowest Y value, which is fixed. Therefore, it is convenient to concentrate on studying the dynamics of these average quantities by using a simplified model of masses (representing entire crystallographic planes) connected by elastic springs.
	
	In this simplified model, we ignore the possible existence of damping forces, usually represented as forces opposing the motion of particles and proportional to their velocity. Hence, for a single particle, the following equation of motion may be used:
	
	\begin{equation}
		\ddot{u} +\Omega^{2}u=0,~~~\Omega=\sqrt {k/m},
		\label{eq:force1}
	\end{equation}
	
	where $k, m$ are the elastic constant of spring and the average mass of a point/plane
	and $\Omega$ is the angular frequency of a harmonic oscillator.
	
	In the case of a chain of connected masses, we may instead write Eq. \ref{eq:force1} as a discrete difference equation:
	
	\begin{equation}
		\ddot{u}_n +2\Omega^{2}u_{n} -\Omega^{2}u_{n+1}-\Omega^{2}u_{n-1}=0.
		\label{eq:force3}
	\end{equation}
	
	Solutions to equations like \ref{eq:force1}, \ref{eq:force3} are well known and form the basis of classical mechanics and physics \cite{Kittel}, \cite{Fetter}. 
	In particular, a broad class of solutions can be written in a form of arbitrary linear combinations of traveling waves, $u_n= A_n \cdot \exp(\rm{ikn}a-\rm{i}\omega t)$, where allowed values of the wave-vector $k$ are determined by boundary conditions, and $a$ is the distance between masses. For instance, when assume periodicity in space dimension ($u_{N+1}=u_{1}$), one arrives at the dispersion relation:

	\begin{equation}
	\omega_n(k_n)=2\Omega \bigg| {\sin\left(\frac{k_n a}{2}\right)} \bigg|,~~~k_n=n\pi/((N+1)a),~~~ 	 n=1,2,3...N-1,
	\label{eq:force37}
	\end{equation}
	
	and the group velocity for small values of k is found independent of k, $c=d\omega_n/dk_n=\Omega a$.
	
	The available literature lacks, however, discussion of the problem of response of a discrete medium to some special boundary conditions relevant to MD simulations, where dynamics of a finite number of particles is studied. Solutions in the form of traveling waves with periodic boundary conditions are obtained when assuming that factoring of time- and space-dependent variables can be carried out. 
	
	The significance of the problem we are dealing with has been first recognized by Erwin Schr\"odinger in 1914 \cite{Erwin}, \cite{Muhlich} and six decades later (in 1974), independently, by de Pater \cite{Pater}. Their works remained largely unknown, and to the best of our knowledge, it was our work \cite{Koziol0}, where equations of de Pater were used for the first time in the analysis of dynamics of stress penetration to the interior of crystals.
	
	We will follow the approach of de Pater, who used the Laplace transform technique to find out the response of a CSM system to a Heaviside-type, an abrupt force applied. The derivation shown here is a shortened repetition of equations of de Pater. We consider it worth repeating that description since the original work of de Pater is probably not broadly available. 
	
	\subsection{De Pater's solution of Heaviside problem by Laplace transform.}
	
	After the Laplace transformation, $\mathcal{L}[u_n](s)=\int_0^\infty u_n(t)\exp(-st)dt$, 
	Eq. \ref{eq:force3} can be written as:
	
	\begin{equation}
		\left( s^2 + 2\Omega^{2} \right) \cdot u^*_{n} -\Omega^{2}u^*_{n+1}-\Omega^{2}u^*_{n-1}=0.
		\label{eq:Laplace01}
	\end{equation}
	
	Let us define variable $z=s/2\Omega$:
	
	\begin{equation}
		2\cdot ( 2z^2  + 1) \cdot u^*_{n} -u^*_{n+1}-u^*_{n-1}=0.
		\label{eq:Laplace02}
	\end{equation}
	
	Let us show that the following is a solution to Eq. \ref{eq:Laplace02}:
	
	\begin{equation}
		u^*_{n} = B e^{-2\alpha(n-1)},~~~{Re}(\alpha)>0.
		\label{eq:Laplace03}
	\end{equation}
	
	After substituting Eq. \ref{eq:Laplace03} to Eq. \ref{eq:Laplace02}, we have:
	
	\begin{equation}
		2\cdot ( 2z^2 + 1) \cdot B e^{-2\alpha(n-1)} -B e^{-2\alpha(n)}-B e^{-2\alpha(n-2)}=0,
		\label{eq:Laplace04}
	\end{equation}
	
	After some lengthy algebraic manipulations we arrive to the following relationship between variables $z$ and $\alpha$, fulfilling
	the condition \ref{eq:Laplace03}: $\sinh(\alpha) = z$.
	
	Parameter $B$ in Eq. \ref{eq:Laplace03} must be determined from the boundary condition:
	
	\begin{equation}
		\ddot{u}_1 +\Omega^{2} (u_{1} -u_{2})=F_1(t)/m,
		\label{eq:Laplace20}
	\end{equation}
	
	where $F_1(t)$ is force acting on the first mass in the chain.
	When a Heaviside-type force is applied, let us write it as $F_1 \cdot H(t)$, where $H(t)$ is the unit Heaviside function. 
	
	Applying Laplace transformation to Eq. \ref{eq:Laplace20} gives us:
	
	\begin{equation}
		s^2\cdot  u^*_1  +\Omega^{2} (u^*_1 -u^*_2)= \mathcal{L}[F_1(t)/m],
		\label{eq:Laplace21}
	\end{equation}
	
	where $\mathcal{L}[F_1(t)/m] = F_1/(m \cdot s)$ is for Heaviside force.
	
	The above can be written as:
	
	\begin{equation}
		\left( s^2 +  \Omega^{2} \right) u^*_1 - \Omega^{2} \cdot u^*_2=\mathcal{L}[F_1(t)/m].
		\label{eq:Laplace22}
	\end{equation}
	
	Using the relations $s=2z\Omega$, $z=\sinh \alpha$, and $u^*_{n} = B e^{-2\alpha(n-1)}$, we finally arrive to:
	
	\begin{equation}
		B =\frac{-\mathcal{L}[F_1(t)]}{2m \cdot z \Omega^2 e^{\alpha}}.
		\label{eq:Laplace25}
	\end{equation}
	
	Let us use superscripts $^H$ and $^L$ to distinguish the Heaviside and linear-in-time boundary cases (the last one to be treated in the next subsection).
	
	Hence,  for $u^{H*}_n$ we have:
	
	\begin{equation}
		u^{H*}_{n} = \frac{F_1}{4m \cdot z^2 \Omega^3 } e^{-\alpha(2n-1)}.
		\label{eq:Laplace30A}
	\end{equation}

	Using the inverse theorem, we can write Eq. \ref{eq:Laplace30A} as follows:
	
	\begin{equation}
		u^H_{n}(t) = \frac{F_1}{4m \Omega^3 } \frac{1}{2\pi j} \int _{\sigma-j\infty}^{\sigma +j\infty} \frac{e^{st} ds}{ z^2 \cdot e^{\alpha(2n-1)}}.
		\label{eq:Laplace31}
	\end{equation}
	
	Defining $\theta = \Omega t$, the above assumes this form:
	
	\begin{equation}
		u^H_{n}(t) = \frac{F_1}{2m \Omega^2 } \frac{1}{2\pi j} \int _{\sigma-j\infty}^{\sigma +j\infty} \frac{e^{2\theta z} dz}{ z^2 \cdot e^{\alpha(2n-1)}}.
		\label{eq:Laplace31A}
	\end{equation}
	
	Velocity $v^H_{n}(t) = \dot{u}_{n}(t)$ differs from $u_{n}(t)$ by a factor $2\Omega z$ under the integral, while spring force $f_n$ is defined through the difference between displacements $u_{n+1}-u_{n}$:
	
	\begin{equation}
		v^H_{n}(t) = \frac{F_1}{m \Omega } \frac{1}{2\pi j} \int _{\sigma-j\infty}^{\sigma +j\infty} \frac{e^{2\theta z} dz}{ z \cdot e^{\alpha(2n-1)}}, ~~~ f^H_{n}(t) = -\frac{F_1}{2\pi j} \int _{\sigma-j\infty}^{\sigma +j\infty} \frac{e^{2\theta z} dz}{ z \cdot e^{2n\alpha}}.
		\label{eq:Laplace31B}
	\end{equation}
	
	Pater uses transformation $\zeta=e^{\alpha}$, i.e. $z=\frac{1}{2}(\zeta -1/\zeta)$, and uses also the generating function for the Bessel coefficients $J_k$:
	
	\begin{equation}
		e^{\theta(\zeta-1/\zeta)} = \sum_{k=-\infty}^{\infty} \zeta ^k J_k (2\theta).
		\label{eq:Laplace35}.
	\end{equation}
	
	Hence, equations on $u_n$, $v_n$ and $f_n$ can be written as follows:
	
	\begin{equation}
		u^H_{n}(t) = \frac{F_1}{m \Omega^2 } \sum_{k=-\infty}^{\infty} J_k (2\theta) \left[ \frac{1}{2\pi j} \oint _{C} \frac{\zeta ^k }{ \zeta^{2n-1}} \frac{(\zeta^2+1)d\zeta}{(\zeta^2-1)^2} \right],
		\label{eq:Laplace36}
	\end{equation}

	\begin{equation}
		v^H_{n}(t) = \frac{F_1}{m \Omega } \sum_{k=-\infty}^{\infty} J_k (2\theta) \left[ \frac{1}{2\pi j} \oint _{C} \frac{\zeta ^k }{ \zeta^{2n}} \frac{(\zeta^2+1)d\zeta}{(\zeta^2-1)} \right],
		\label{eq:Laplace37}
	\end{equation}

	\begin{equation}
		f^H_{n}(t) = -F_1 \sum_{k=-\infty}^{\infty} J_k (2\theta) \left[ \frac{1}{2\pi j} \oint _{C} \frac{\zeta ^k }{ \zeta^{2n+1}} \frac{(\zeta^2+1)d\zeta}{(\zeta^2-1)} \right].
		\label{eq:Laplace38}
	\end{equation}

	Now, the problem reduces to computing integrals of the kind:
	
	\begin{equation}
		\frac{1}{2\pi j} \oint \frac{\zeta ^l d\zeta}{(\zeta^2-1)},~~~\frac{1}{2\pi j} \oint \frac{\zeta ^l d\zeta}{(\zeta^2-1)^2}.
		\label{eq:Laplace38A}
	\end{equation}

	Pater finds out that for any $l<0$, and for any even $l>0$, the integrals are equal to 0. For positive odd values of $l$, $l=1,3,5...$ the first integral is equal to 1, while the second is $(l-1)/2$.
	
	The above allows us, after some algebraic manipulations, to arrive finally to the following expressions on quantities with Heaviside boundary condition:
	
	\begin{equation}
		u^H_n(t) = \frac{F_1}{m \Omega^2 } \left[J_{2n}(2\theta) + \sum _{k=2n+2,2n+4,...} ^{\infty} (k-2n+1) J_k(2\theta)\right],
		\label{eq:Hun}
	\end{equation}
	
	\begin{equation}
		v^H_n(t) = \frac{F_1}{m \Omega }\left[J_{2n-1}(2\theta) + 2\cdot \sum _{k=2n+1,2n+3,...} ^{\infty} J_k(2\theta)\right],
		\label{eq:Hvn}
	\end{equation}
	
	\begin{equation}
		f^H_n(t) = -F_1\left[ 1-J_0(2\theta) -2\cdot \sum _{k=2,4,...} ^{2n} J_k(2\theta) + J_{2n}(2\theta)\right],
		\label{eq:Hfn}
	\end{equation}

	where $\theta=\Omega t$.

	\subsection{The linear-in-time surface pressure.}
	
The other case which is important in MD simulations is when a linear-in-time force is applied to the surface. Let us write it as $F_1(t) =h\cdot t$. 
The Laplace transform of that function is: $\mathcal{L}[F_1(t)/m]=h/(m \cdot s^2)$. Therefore, instead of Eq. \ref{eq:Laplace30A}, we have now:

	\begin{equation}
		u^{L*}_{n} = \frac{h}{8m \cdot z^3 \Omega^4 } e^{-\alpha(2n-1)}. 
	\label{eq:L00}
	\end{equation}

In result, the inverse transformation of $u^{L*}_{n}$ gives us:

	\begin{equation}
		u^L_{n}(t) = \frac{h}{4m \Omega^3 } \frac{1}{2\pi j} \int _{\sigma-j\infty}^{\sigma +j\infty} \frac{e^{2\theta z} dz}{ z^3 \cdot e^{\alpha(2n-1)}}.
	\label{eq:L01}
	\end{equation}

Since velocity $v^L_{n}(t) = \dot{u}^L_{n}(t)$ differs from $u^L_{n}(t)$ by a factor $2\Omega z$ under the integral, we have:

	\begin{equation}
		v^L_{n}(t) = \frac{h}{2m \Omega^2 } \frac{1}{2\pi j} \int _{\sigma-j\infty}^{\sigma +j\infty} \frac{e^{2\theta z} dz}{ z^2 \cdot e^{\alpha(2n-1)}}
	\label{eq:L02}
	\end{equation}

	We observe that $u^H_n$ and $v^L_n$, as given by Eqs. \ref{eq:Laplace31A} and \ref{eq:L02}, respectively, differ only by a multiplication factor. Therefore, we can write right away:
	
	\begin{equation}
		v^L_n(t) = \frac{h}{m \Omega^2 } \left[J_{2n}(2\theta) + \sum _{k=2n+2,2n+4,...} ^{\infty} (k-2n+1) J_k(2\theta)\right].
		\label{eq:Lvn}
	\end{equation}
	
	Let us use the fact that $v^L_n(t)$ is a time derivative of $u^L_n(t)$. We assume the last quantity to be given by an infinite sum:
	
	\begin{equation}
		u^L_n(t) = \frac{h}{m \Omega^3 } \sum_{k=0}^{\infty} a_{n,k} \cdot J_k (2\theta)
		\label{eq:Lun0}
	\end{equation}
	
	Since $\dot{J}_{\nu} (2\theta)=\Omega(J_{\nu-1}(2\theta)-J_{\nu+1}(2\theta))$, after differentiating $u^L_n(t)$ and comparing coefficients of expansion of $\dot{u}^{L}_n(t)$ with these in Eq. \ref{eq:Lvn},  we obtain a recursive
	relation: 
	
	\begin{equation}
		a_{n,k+2}= (k-2n+1) +a_{n,k}, 
		\label{eq:Lun1}
	\end{equation}
	
	with $a_{n,i}=0$ for $i<2n+1$, and $a_{n,i}=0$ for any even $i$. It is easy to check by mathematical induction that coefficients $a_{n,k}$ can be written as:
	
	\begin{equation}
		a_{n,k}= (k-2n+1)^2/4.
		\label{eq:Lun2}
	\end{equation}
	
	Hence, we can write:
	
	\begin{equation}
		u^L_n(t) = \frac{h}{m \Omega^3 } \sum_{k=2n+1,2n+3...}^{\infty} (k-2n+1)^2/4 \cdot J_k (2\theta)
		\label{eq:Lun3}
	\end{equation}
	
	The spring force $f^L_n$ is related to the difference between $u^L_{n+1}$ and $u^L_{n}$. Therefore, using Eq. \ref{eq:Lun2}, we find coefficients for $J_k$ from the difference $a_{n+1,k}-a_{n,k}$. The result can be written as:
	
	\begin{equation}
		f^L_n(t) = -\frac{h}{\Omega}\sum_{k=2n+1,2n+3...}^{\infty} (k-2n) \cdot J_k (2\theta).
		\label{eq:Lfn4}
	\end{equation}
	
	Equations \ref{eq:Lun3}, \ref{eq:Lvn} and \ref{eq:Lfn4} form a full set of solutions on $u^L_n$, $v^L_n$, and $f^L_n$.

\section*{References}

\bibliographystyle{iopart-num}
	\bibliography{Analytical3}
\label{Bibliography}

\end{document}